# Hydrodynamics, Vortices and Angular Momenta of Celestial Objects

*C Sivaram*

Indian Institute of Astrophysics, Bangalore, 560 034, India

Telephone: +91-80-2553 0672; Fax: +91-80-2553 4043

e-mail: sivaram@iiap.res.in

*Kenath Arun*

Christ Junior College, Bangalore, 560 029, India

Telephone: +91-80-4012 9292; Fax: +91-80- 4012 9222

e-mail: kenath.arun@cjc.christcollege.edu

**Abstract:** The current observational evidences suggest there are about hundred billion galaxies in the observable universe and within each, on an average, about hundred billion stars. But no cosmological model indicates as to why there are these many galaxies and stars. In this paper we invoke the property of non-irrotational hydrodynamic flow in order to explain how a primordial rotation (as considered in a recent paper) of the universe broken up into vortex line structures, can indeed lead to formation of a large number of galactic structures and these in turn can lead to equally large number of stars within each galaxy.

In a recent paper (Sivaram & Arun, 2011a) a cosmological model involving the primordial rotation of the universe was invoked to understand the origin of the rotation or spin of objects over a wide range of masses from stars to galaxies. However this did not address the question of the distribution of the primordial angular momentum over the very large number of these objects. For instance, there are about a hundred billion stars in a typical large galaxy and a hundred billion galaxies in the universe.

It is clear that hydrodynamics and turbulence have to be invoked. Some models of galaxy formation have explicitly included those effects (Krishan & Sivaram, 1991; Bournaud, 2010). Again we have models, where cosmic strings have played a role as 'seeds' to trigger galaxy formation (Sivaram, 1987; Manías, et al, 1986). The study of cosmic strings from the point of view of defects in condensed matter physics, has led to experiments in analogy with superfluids, wherein rotating a container of superfluid leads to the angular momentum being distributed in an array of vortices with quantized circulation (Finne et al, 2006). For early work on relevance of superfluid vacuum state to phase transitions in the early universe see (Volovik, 1984)

In what follows, we shall invoke this property of non-irrotational hydrodynamic flow, to understand how the primordial rotation of the universe (as suggested in (Sivaram & Arun, 2011a) and earlier works) broken up into vortex line structures, can indeed lead to formation of a large number of galactic structures, which in turn can lead to equally large number of stellar objects (within a galaxy).

For a fluid with non-irrotational flow,

$$\nabla \times v_s \neq 0 \qquad \ldots (1)$$

In the case of a superfluid, equation (1) holds at singular line core of a quantised vortex line, and the quantum of circulation is governed by the Onsager-Feynman relation:

$$\oint v \, dr = n \frac{\hbar}{m_s} \qquad \ldots (2)$$

$$\oint v \, dr = \iint \nabla \times v_s \, dA \qquad \ldots (2a)$$

Or for an angular velocity $\omega$:

$$\oint \omega r \, dr = n \frac{\hbar}{m_s} \qquad \ldots (3)$$

($m_s$ is the mass of the particle)

For a rotating container, we have: $2\pi r.v = n\hbar/m_s$, giving the number of vortices per unit area as $2\omega m_s/\hbar$, and the total number of vortices as:

$$N = \frac{\omega.2\pi r^2 m}{\hbar} \quad \text{... (4)}$$

In actual experiments, for a container rotating at 1000rps, about $2\times10^6$ vortices/cm$^2$ were formed in agreement with the above formulae.

Following the numbers given in (Sivaram & Arun, 2011a), we can estimate the number of galaxy mass 'vortices' in a region (container) of the size of the universe ($R = 10^{28} cm$) with a primordial angular velocity $\omega = 2\times10^{-18} rad/s$. $\hbar$ would be replaced by the typical galaxy angular momentum of $\sim 10^{74} ergs.s$ (as noted in (Sivaram, C. & Arun, K. 2011b)).

This gives the number of galaxies which would have been formed as:

$$N_{gal} = \frac{2\times10^{-18} \times 10^{57} \times 10^{45}}{10^{74}} \approx 10^{11} \quad \text{... (5)}$$

(we have used for a typical vortex mass, a galaxy mass, $m \sim 10^{45} g$, $R^2 \sim (3\times10^{28} cm)^2 \approx 10^{57} cm^2$)

This is the actual number of galaxies in the universe. For an individual galaxy, the number of substructures (stars in this case), is again given using equation (4) but now, with $m \sim 10^{33} g$, $R \approx 10^{23} cm$ (typical galaxy size region), $J \approx 10^{52} erg.s$ (typical for a star):

$$N_{stars} = \frac{2\times10^{-18} \times 10^{46} \times 10^{33}}{10^{52}} \approx 10^{11} \quad \text{... (6)}$$

Thus typically $\sim 10^{11}$ stars form in individual galaxies. So apart from explaining the observed angular momenta of galaxies and stars, from a universe in primordial rotation, we are also able to obtain the numbers of galaxies and stars by using the hydrodynamic vortex analogy for which there is strong experimental laboratory evidence (i.e. for equations (3) and (4)).